# INFORMATION FILTERING BASED ON WIKI INDEX DATABASE

ALEXANDER V. SMIRNOV
ANDREW A. KRIZHANOVSKY

*St. Petersburg Institute for Informatics and Automation
of the Russian Academy of Sciences (SPIIRAS),
St.Petersburg, 199178, Russia
Email: {smir, aka}@iias.spb.su*

In this paper we present a profile-based approach to information filtering by an analysis of the content of text documents. The Wikipedia index database is created and used to automatically generate the user profile from the user's document collection. The problem-oriented Wikipedia subcorpora are created (using knowledge extracted from the user profile) for each topic of user interests. The index databases of these subcorpora are applied to filtering information flow (e.g., mails, news). Thus, the analyzed texts are classified into several topics explicitly presented in the user profile. The paper concentrates on the indexing part of the approach. The architecture of an application implementing the Wikipedia indexing is described. The indexing method is evaluated using the Russian and Simple English Wikipedia.

## 1. INTRODUCTION

The goal of an information filtering system is to alleviate the work of user, to make more effective the persistent search of relevant information. A software module for text filtering is the important part of recommender systems and information filtering systems. Recommender systems could be classified as content-based systems (presented in this work) and collaborative filtering systems.[7]

The recommender system could be based on thesaurus (e.g., WordNet [11]) or an ontology.[12] The experimental comparison [2, 8, 19] of algorithms *searching for related terms* based on WordNet [1, 5, 10, 15-16, 20], GermaNet [14] and English Wikipedia [19] shows an advantage of Wikipedia.

The using of Wikipedia (WP) provides the following benefits:

- largeness of size; WordNet 3.0 contains 155 thousands of words and 118 thousands of synsets, but English Wikipedia contains 946 million of words[a] and 176 thousand of categories.[b] At the same time, the huge size is one of the difficulties that arise when implementing a search algorithm. Thus, the Simple English Wikipedia[c] is usually used for an initial algorithm testing, since it is about 100 times smaller than English Wikipedia (see, e.g., experiments in [17]).
- WordNet gloss (a brief explanatory note) is shorter than an WP article[d]. It is important for a full-text analysis, e.g., the precision of search in WP [19] has surpassed the precision of the Adapted Lesk algorithm which is based on the glosses analysis.[1] In the paper [2] the algorithm of artificial increasing the size of glosses ("extended gloss overlaps") was proposed to cope with the brevity of glosses.
- Wikipedias are presented in many languages, not only in English.

The development of the text filtering approach based on the wiki indexing requires: (i) to develop the text filtering approach, (ii) to design the architecture of the wiki indexing system, (iii) to implement the indexing system and run the experiments. The paper structure corresponds to the formulated tasks.

---

[a] As of 27 January 2008, see http://en.wikipedia.org/wiki/Wikipedia:Size_comparisons.
[b] As of 30 October 2006, see http://stats.wikimedia.org/EN/TablesWikipediaEN.htm.
[c] See http://simple.wikipedia.org.
[d] The average number of words per article is 400, as of October 2005, see http://en.wikipedia.org/wiki/Wikipedia:Words_per_article.

## 2. THE TEXT FILTERING APPROACH BASED ON THE WIKIPEDIA INDEX DATABASE

The approach to information filtering consists of four stages (Fig. 1):

I. *Wikipedia Indexing*.

II. *Profile Generation*. At this moment we have (i) the index WP database (thanks to the step "*1. Indexing*", Fig. 1), and (ii) the collection of documents (provided by the user) which implicitly presents a set of topics of user interests. It is needed:

1. Extract a list of lemmas (terms) for each user document with the help of GATE [4] and Lemmatizer [18];
2. Find similar WP articles (similar to the user documents) by ESA algorithm;[6]
3. Get *a list of WP categories* for each article with the help of Synarcher API.[9]

The list of WP categories (which could be very large) should be divided to groups ("*4. Clustering*"). The goal is to get the set of subsets of categories so that each subset would correspond to one topic in the profile. Categories in one subset should not include each other (when a category is a parent of another one). At this stage user can *check profile* and add / remove categories in order to refine his own topics.

III. *Problem-Oriented Index DB Creation*. For each subset of categories the texts of WP articles are extracted from the WP database. Only those articles are extracted, which belong to these categories or their subcategories ("*5. Population*"). The created problem-oriented subcorpus of WP is used to create the problem-oriented index DB ("*6. Indexing*").

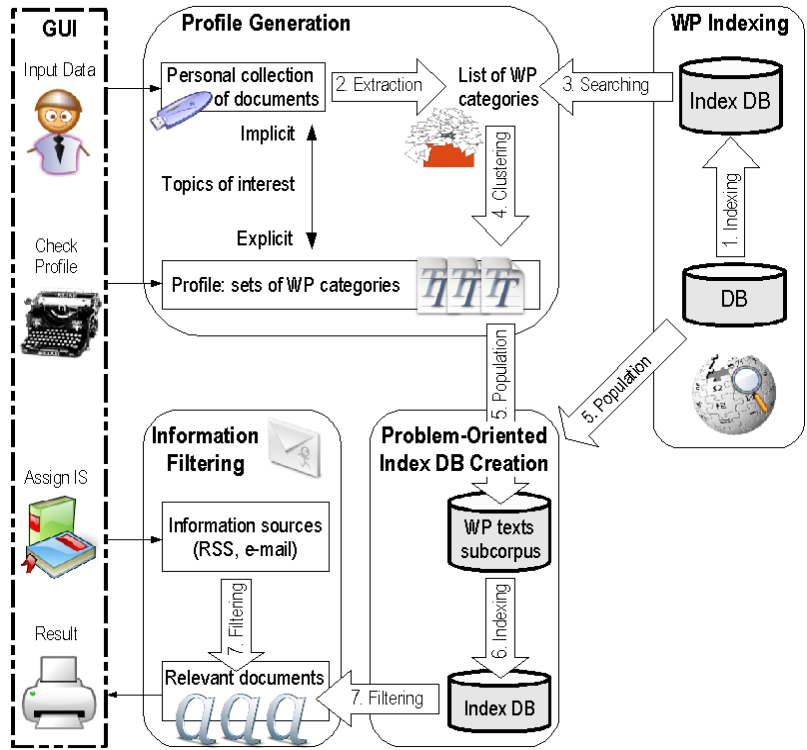

Fig. 1. **The text filtering approach based on the Wikipedia (WP) index database.**

IV. *Information Filtering*. The user selects RSS-sites, defines e-mail parameters and denotes them as information sources (IS) for the recommender system. For each IS the user selects the corresponding topic from his profile, i.e. the user defines the related problem-oriented index DB for each IS. If new texts appear in an IS then it is needed:
1. Extract a list of lemmas (terms) for each text;
2. Find these terms in the corresponding problem-oriented index DB, get weights of these terms from DB, calculate the total weight of the text;
3. Present the *result* to the user, i.e. the list of texts ordered by weight.

Thus, we have proceeded under the assumption that there is an available WP index DB and there is an algorithm for creating the index DB from the text corpus. This leads us to discuss this question in more detail.

## 3. ARCHITECTURE OF WIKI INDEXING SYSTEM

The wiki indexing system requires three groups of input parameters (Fig. 2).

1. The *Language* that defines the language of Wikipedia (one of 254 as of 16 Jan 2008) and the language of lemmatizing.[e] The language of WP should be defined for the correct extracting texts from texts in wiki-format (see Fig. 2, function "Convert wiki-format to text" of the software module "Wikipedia Application").
2. *Database location* that is a set of parameters (host, port, login, password) for connecting to the remote DB (WP and index).
3. *TF-IDF constraints* that define the size of the result index DB.

*Control Application* (Fig. 2) performs the following three steps for each article. (1) The article is extracted from WP database and transformed into the text in a natural language. (2) GATE[4] and Lemmatizer[18] (integrated by the interface application RussianPOSTagger)[f] generate the list of lemmas and their frequencies in the text. (3) The following data are stored to the index DB: (i) the lemmas, (ii) the relation between the lemma and the WP article, (iii) the frequencies of these lemmas in the corpus, which are incremented.[g]

---

[e] Since Lemmatizer has three internal DB (for Russian, English, and German).
[f] See more information about Lemmatizer and RussianPOSTagger at http://rupostagger.sourceforge.net.
[g] The created index database for Russian Wikipedia and Simple English Wikipedia are available at: http://rupostagger.sourceforge.net, see packages *idfruwiki* and *idfsimplewiki*.

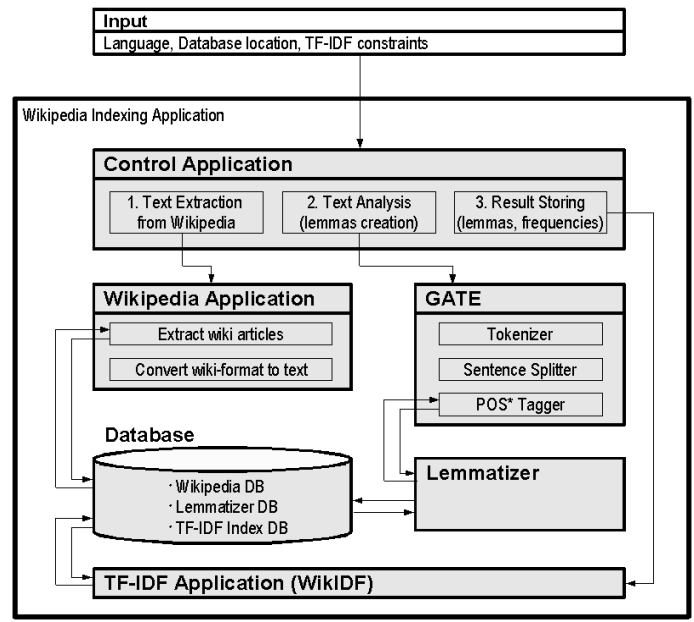

Fig. 2. **Wiki indexing system architecture (*POS means Part Of Speech*).**

It should be noted that two functions of the module "*Wikipedia Application*" (Fig. 2) and API for access to "*TF-IDF Index DB*" were realized in the program Synarcher.[h] The defining of input parameters and start of indexing are implemented in the Synarcher module: WikIDF.[i]

---

[h] This functionality is available in *Synarcher* (from version 0.12.5), see http://synarcher.sourceforge.net.
[i] WikIDF is a console application, which depends on the Java Synarcher package. WikIDF is bundled with Synarcher.

## 4. CONCLUSIONS

The development of an application for indexing wiki-texts has been started. Some initial experiments to build the index DB of Simple English (14/02/2008) and Russian Wikipedia (20/02/2008) have been carried out. The statistical data of the result index DB are presented in Table 1.

Table 1. **Statistics of the index Wikipedia databases.**

| Wikipedia database | Simple English (SE) | Russian (R) | R / SE |
|---|---|---|---|
| Wordform (unique words in corpus), million | 0.149 | 1.43 | 9.6 |
| Word-article relations (<1000 for the word), million | 1.65 | 15.71 | 9.5 |
| Words in corpus, million | 2.28 | 32.93 | 14.4 |
| Size of dump file (archived) of index DB, MB | 7.15 | 77.5 | 10.8 |

## Acknowledgments

This work is supported in part by the grant # 08-07-00264 of the Russian Foundation for Basic Research and projects supported by the Russian Academy of Sciences (RAS) # 14.2.35 and # 1.9.